\documentclass[aps,prl,twocolumn,letterpaper,showpacs,preprintnumbers,amsmath,amssymb]{revtex4}

\usepackage{epsfig,amssymb}
\usepackage{graphicx}
\usepackage{dcolumn}
\usepackage{bm}
\usepackage{dcolumn} 
\usepackage{multirow} 

\begin{document}

\title{Classical density-functional theory of inhomogeneous water
including explicit molecular structure and nonlinear dielectric
response}

\author{Johannes~Lischner}
\author{T.A.~Arias} 
\affiliation{Laboratory of Atomic and Solid State Physics, Cornell University, Ithaca, New
York 14853} 
\date{\today}

\begin{abstract}
We present an accurate free-energy functional for liquid water written
in terms of a set of effective potential fields in which fictitious
noninteracting water molecules move. The functional contains an
\emph{exact} expression of the entropy of noninteracting molecules and
thus provides an ideal starting point for the inclusion of complex
inter-molecular interactions which depend on the \emph{orientation} of
the interacting molecules.  We show how an excess free-energy
functional can be constructed to reproduce the following properties of
water: the dielectric response; the experimental site-site correlation
functions; the surface tension; the bulk modulus of the liquid and the
variation of this modulus with pressure; the density of the liquid and
the vapor phase; and liquid-vapor coexistence.  As a demonstration, we
present results for the application of this theory to the behavior of
liquid water in a parallel plate capacitor.  In particular, we make
predictions for the dielectric response of water in the nonlinear
regime, finding excellent agreement with known data.

\end{abstract}
\maketitle

\section{Introduction}
Water is the most important liquid on earth. Although its uniform
phase already exhibits many fascinating properties and is currently an
area of intensive research, the importance of the \emph{inhomogeneous}
phase in the chemical and biological sciences is even greater: as a
solvent, water is ubiquitous in physical chemistry, biochemistry and
electrochemistry.  The associated phenomena include hydrophobic
interactions \cite{ChandlerHydrophobic, HydrophobicReview}, protein
folding \cite{ProteinBerne,ProteinTarek}, the electrochemical
interface \cite{DiffusionElChemInt}, and phase transitions of confined
liquids \cite{ConfinedHummer,ConfinedLeng}.  Due to the complex
interplay of hydrogen bonding, long-range polar interactions, and
short-range excluded volume effects, developing a tractable theory to
describe the solvent in the aforementioned systems remains a challenge
\cite{SimpleLiquids}.

Despite the importance, inherent interest, and extensive experimental
study of water, most existing theories of the inhomogeneous phase of
water either lack the accuracy to describe the above phenomena in a
quantitatively satisfying way or become computationally prohibitive.
The latter is particularly true for applications to complex systems,
such as dissolved biomolecules or electrochemical interfaces.  Simple
dielectric continuum theories \cite{TomasiCon1,TomasiCon2} capture the
dielectric response of the solvent at long length scales, but fail to
describe more subtle effects at interfaces, such as the reorganization
of the hydrogen-bond network or the binding of water molecules to
solutes. Atomistic theories, on the other hand, describe such local
effects quite well, but require the explicit treatment of many solvent
molecules and also demand long simulation times to calculate
statistically reliable thermodynamic averages.  Since the full-fledged
\emph{ab initio} molecular dynamics description of the solvent is only
feasible for the smallest solutes, generally additional approximations
become necessary in practice: Bernholc and coworkers, for example,
have introduced a clever scheme which employs the full \emph{ab
  initio} description only for a few solvent molecules adjacent to the
solute and use rigid geometries and frozen electron densities for the
remaining molecules \cite{Bernholc}.  In other studies, combination of
\emph{ab initio} methods for the solute with classical force field
approaches \cite{LevitQMMM,KollmannQMMM,KarplusQMMM} for the solvent
yields additional simplifications.  Despite the simplifications which
such hybrid approaches allow, the need to compute many configurations
to sample phase-space properly ultimately limits the size of systems
which can be studied and thus the applicability of such approaches.

Ideally, one would have a rigorous description of the solvent which is
able to combine the advantages of continuum theories (large system
sizes and {\em implicit} thermodynamic averaging) with an explicit
geometric description of the solvent molecule.  Here, we show that
just such a theory can be developed for liquid water by starting with
the quantum mechanical free-energy functional for both the electrons
and nuclei which comprise the liquid and, by ``integrating out'' the
electrons, proceeding to construct a density functional in terms of
atomic site densities alone.  Such ``classical'' density-functional
theories, which are founded on a number of exact theorems
\cite{Mermin,Ashcroft}, have been applied successfully to the study of
simple liquids in the past \cite{SaamEbner,AshcroftCurtin2}.
Historically, application of this method begins with a hard sphere
reference system that is usually augmented by terms that capture weak
long-range attractive forces \cite{Weeks}. Unfortunately, a hard
sphere reference system is a poor starting point for the description
of water because of the strong anisotropic short-range interactions
arising from the molecular structure, including effects such as
hydrogen bonding.

To remedy this, Chandler and coworkers \cite{Chandler1, Chandler2,
  ChandlerWater} introduced a density-functional theory for general
molecular liquids in terms of a set of densities, one for each
``interaction site'' on the molecule (typically atomic centers).
Although their theory successfully predicted the correct
hydrogen-bonded structure of ice \cite{ChandlerWater}, it suffered
from a significant flaw: to make calculations tractable, the ``united
atom approximation'', which takes all of the sites on the molecule to
sit at a single point, had to be invoked.  This severe
oversimplification of the molecular structure prohibited the
description of dielectric properties and stalled the further
development of this promising approach. In \cite{JLTAA}, we eliminated
the need for the united atom approximation by writing the free energy
of a general molecular liquid as a functional of a set of effective
potentials in which fictitious non-interacting molecules move. This
advance allows for the exact evaluation of the entropy associated with
the geometric structure of the molecules in a numerically feasible way
and forms the basis for the construction of a new class of accurate
density functionals for fully interacting {\em molecular} liquids.

In this work, we apply our general theory of molecular liquids to
water and demonstrate how an accurate free-energy functional can be
developed to reproduce the equation of state, the experimental
site-site correlation functions, and the macroscopic dielectric
response.  The organization of the work is as follows. Section 2 gives
a brief review of the general theory of molecular liquids introduced
in \cite{JLTAA}.  Section 3 discusses the application of this approach
to the particular case of water and introduces new computational
techniques to simplify calculations of triatomic liquids. Section 4
presents results obtained using this new microscopically-informed
continuum theory to study the behavior of water in a parallel plate
capacitor.  Here, we explore the differences between capacitor plates
consisting either of purely repulsive hard walls or of walls with a
local region of attraction.  We also explore the dielectric response
of water in both the linear and the nonlinear regimes of the applied
field.  Comparing our results to previous works based on explicit
molecular simulations, we find quite encouraging agreement. Section 5
then concludes this work.

\section{Density-functional theory of molecular liquids}

\subsection{Kohn-Sham Approach and Entropy Functional}

In \cite{JLTAA}, we showed that the grand free energy $\Omega$ of an
assembly of interacting identical molecules (consisting of $M$ atoms
or ``interaction sites'' each) can be expressed as a functional of a
set of effective potentials
$\mathbf{\Psi}=\{\Psi_{1}(\bm{r}),...,\Psi_{M}(\bm{r})\}$, in which a
corresponding set of fictitious non-interacting molecules move.
Specifically, we wrote
\begin{eqnarray}
\Omega[\mathbf{\Psi}]= \Omega^{(ni)}[\mathbf{\Psi}]-
\sum_{\alpha=1}^{M}\int d^{3}r \left(
\Psi_{\alpha}(\mathbf{r})-\phi_{\alpha}(\mathbf{r})+\mu_{\alpha}\right)
n_{\alpha}(\mathbf{r}) \nonumber\\
+U[\mathbf{n}].
\label{FullOmega}
\end{eqnarray}
Here, $\phi_{\alpha}(\mathbf{r})$ is the site-dependent external
potential (with $\alpha$ referring to the different atomic sites on
each molecule), $\mu_{\alpha}$ is a site specific chemical potential and
$U[\mathbf{n}]$ denotes the excess free energy due to inter-molecular
interactions, which is a functional of the set of site densities
$\bm{n}=\{n_{1}(\bm{r}),...,n_{M}(\bm{r})\}$.  Finally,
$\Omega^{(ni)}[\mathbf{\Psi}]$ is the grand free energy of an assembly
of noninteracting molecules in a set of effective potentials
$\mathbf{\Psi}$.  This free energy is known explicitly and is {\em exactly}
\begin{eqnarray}
\Omega^{(ni)}[\mathbf{\Psi}]=-k_{B}Tn_{l} 
\int d^{3M}r\;s(\{\mathbf{r}_\alpha\}) 
e^{-\beta\sum_{\alpha=1}^{M}\Psi_{\alpha}(\mathbf{r}_{\alpha})}
\label{freeEni},
\end{eqnarray}
where $n_{l}$ is the reference density at vanishing chemical
potentials (which we will later take to be the bulk liquid density)
and $s(\{\mathbf{r}_\alpha\})$, which describes the geometry of the
molecule, is the intra-molecular distribution function. Note that we
take the densities in (\ref{FullOmega}) to be themselves functionals
of the effective fields via
\begin{equation} \label{eq:getns}
n_{\alpha}(\mathbf{r})=\frac{\delta\Omega^{(ni)}}{\delta\Psi_{\alpha}(\mathbf{r})}[\mathbf{\Psi}].
\end{equation}
Finally, the effective potentials which minimize $\Omega[\mathbf{\Psi}]$
determine the equilibrium site densities and allow for the calculation
of various equilibrium properties of the liquid.
 
The introduction of effective potentials instead of site density
fields as fundamental variables in the density-functional theory of
classical molecular liquids allows for the first time a
computationally tractable approach to treat the non-interacting system
exactly.  This parallels the introduction of Kohn-Sham orbitals as
fundamental variables in the density-functional theory of electronic
structure \cite{KohnSham}. In both cases, the construction of highly
accurate density functionals is made possible by mapping a system of
interacting particles onto a fictitious system of noninteracting
particles with the same equilibrium densities. We mention that our
approach of minimizing the free energy with respect to a set of
\emph{effective potentials} is known in the electronic structure
context as the optimized effective potential method
\cite{TalmanShadwick}.

\subsection{Excess Free Energy Functional}

In \cite{JLTAA}, we also present a recipe for the construction of the
excess free energy functional $U$ in (\ref{FullOmega}). As usual with
density-functional theories, this construction is difficult.  As in
our previous work, we follow the lead of Kohn and Sham and construct a
free-energy functional that reproduces established results for the
homogeneous phase in the limit of vanishing external fields.  Namely,
we expand $U$ in a power series about the uniform liquid, and formally
group all terms except the quadratic part, which is analogous to the
Hartree energy in electronic-structure theory, into an unknown
functional $F^{ex}[\mathbf{n}]$, which is then the analogue to the
electronic exchange-correlation functional, whose form, being unknown,
must ultimately be treated in some approximate way.

To proceed, we separate the kernel of the quadratic part of $U$ into a
long-wavelength part $K_{\alpha\gamma}$, known analytically for the
case of rigid molecules, and a remainder $C_{\alpha\gamma}$, to be
determined from experimental data.  The next paragraph describes
$K_{\alpha\gamma}$ in detail, and the set of functions
$C_{\alpha\gamma}$ will be constructed below such that the resulting
functional reproduces the experimental correlation functions in the
uniform phase.  When these two parts are determined, they combine with
$F^{ex}$ to give $U$ as
\begin{eqnarray}
U[\mathbf{n}]=\frac{1}{2}\sum_{\alpha,\gamma=1}^{M}\int
d^{3}r \int d^{3}r'\; n_{\alpha}(\mathbf{r})\left\{
K_{\alpha\gamma}(\mathbf{r},\mathbf{r'}) \right. \nonumber\\
\left. +C_{\alpha\gamma}(\mathbf{r},\mathbf{r'})\right\} n_{\gamma}(\mathbf{r'})
+F^{ex}[\mathbf{n}].
\label{U}
\end{eqnarray}

The derivation of $K_{\alpha\gamma}$\cite{JLTAA} considers a
collection of rigid neutral molecules with each interaction site
carrying a partial charge $q_{\alpha}$ with {\em no} assumptions
regarding the form of the non-electrostatic part of the interaction
--- the theory neither is perturbative in the interaction strength nor
presupposes a decomposition into pairwise interaction potentials.  In
\cite{JLTAA}, we then prove rigorously that, for a certain class of
liquids including all diatomic liquids and also water, the leading
order term in a long-wavelength expansion of
$\partial^{2}_{\mathbf{n}}U$, the Hessian of U for rigid molecules, is
\begin{equation}
K_{\alpha\gamma}(\mathbf{k})=\left(\frac{\epsilon}{\epsilon-1}-
\frac{\epsilon^{(ni)}}{\epsilon^{(ni)}-1}\right)
\frac{4\pi}{\mathbf{k}^{2}}q_{\alpha}q_{\gamma},
\label{K}
\end{equation}
where $\epsilon$ is the macroscopic dielectric constant and
$\epsilon^{(ni)}$ is the dielectric constant of a system with
intra-molecular correlations only.
Because $K_{\alpha\gamma}$ now captures the singular long-wavelength
features of the response function, $C_{\alpha\gamma}$, which we define
as the difference between $\partial^{2}_{\mathbf{n}}U$ (given by the
experimental correlation functions) and the sum of $K_{\alpha\gamma}$
and $\partial^{2}_{\mathbf{n}}F^{ex}$, must now be smooth near the
origin and thus amenable to numerical approximations. As discussed
below, we define $F^{ex}$ such that its Hessian matches the
constant term in the long-wavelength expansion of
$\partial^{2}_{\mathbf{n}}U$. Therefore, the functions
$C_{\alpha\gamma}$ have to \emph{vanish} for small $\mathbf{k}$.

Next, to approximate $F^{ex}$, we begin by noting that in the case of
zero external fields all densities are equal and the first quadratic
term ($K$) in (\ref{U}) vanishes because of charge neutrality.
Anticipating that, by definition, the matrix function $C$ vanishes in
the long-wavelength limit, $F^{ex}$ then captures all of the internal
energy of the uniform phase (and the long-wavelength limit of the
non-singular part of the Hessian $\partial^{2}_{\mathbf{n}}U$) and can
be expressed as $F^{ex}=Vf^{ex}(n)$, with $V$ being the volume and $n$
the average molecular density.  Due to the presence of multiple
density fields, generalizing the above expression for $F^{ex}$ to the
inhomogeneous case is more difficult than for the analogous
exchange-correlation energy in electronic structure theory. Moreover,
because of the strong correlations induced by excluded volume effects,
{\em purely local} excess functionals fail to describe the liquid
state \cite{CurtinAshcroft}.  We therefore approximate $F^{ex}$ with a
simplified {\em ansatz} in the spirit of weighted density-functional
theory \cite{CurtinAshcroft}, but generalized to multiple species,
\begin{equation}
F^{ex}[\bm{n}]=\int
d^{3}r\sum_{i}p_{i}
f^{ex}\left(\sum_{\gamma=1}^{M}b_{\gamma}^{i}\overline{n}_{\gamma}(\mathbf{r})\right),
\label{Fex}
\end{equation} 
where we introduce the weighted densities
$\overline{n}_{\gamma}(\mathbf{r})=\int d^{3}r' (\pi r_{0}^{2})^{-3/2}
\exp(-|\mathbf{r}-\mathbf{r'}|^{2}/r_{0}^{2})n_{\gamma}(\mathbf{r}')$,
with $r_{0}$ being a ``smoothing'' parameter ultimately
fit to the experimental surface tension.  To reduce to
the correct form in the uniform case, $p_{i}$ and $b^{i}_{\gamma}$ must
fulfill $\sum_{i}p_{i}=1$ and
$\sum_{\gamma=1}^{M}b_{\gamma}^{i}=1$.

For the scalar function $f^{ex}(n)$, we use a polynomial fit to
various bulk thermodynamic properties of the liquid.  The condition
that $C$ vanishes in the long-wavelength limit subsequently fixes
$p_{i}$ and $b^{i}_{\gamma}$. For a given $r_0$, this then completely
specifies our approximation to $F^{ex}$.  Next, relating
$K+C+\partial^2 F^{ex}$ to the density-density correlation functions
through the Ornstein-Zernike relation gives the matrix function $C$
for a given $r_0$.  Finally, we can determine the smoothing parameter
$r_0$ by adjustment until calculations of the liquid-vapor interface
give the correct surface tension.

\section{Application to Water}
\subsection{Intra-molecular Free Energy}

Assuming rigid bonds of fixed length (and thus rigid angles as well)
the intra-molecular distribution function of a water molecule
(normalized to integrate to the volume of the system) is
\begin{eqnarray}
s(\mathbf{r}_{0},\mathbf{r}_{1},\mathbf{r}_{2})=\frac{\delta(|\mathbf{r}_{10}|-B)}{4\pi B^{2}}
\frac{\delta(|\mathbf{r}_{20}|-B)}{4\pi B^{2}}\nonumber\\
\times 2\delta(\hat{\mathbf{r}}_{10}\cdot\hat{\mathbf{r}}_{20}-\cos\theta_{B}),
\label{sWater}
\end{eqnarray}
where $B=1$~\AA~is the oxygen-hydrogen bond length taken from the
SPC/E \cite{BerendsenSPCE} model of water, $\theta_{B}$ is the
bond angle between the hydrogens and taken to be the tetrahedral angle
like in the SPC/E model. Also, the label $0$ refers to the oxygen
atom and the labels $1$ and $2$ to the two hydrogens, and we define
$\mathbf{r}_{ij}\equiv\mathbf{r}_{i}-\mathbf{r}_{j}$.  Contrary to the
case of diatomic molecules, where the assumption of rigid bonds turns
$\Omega^{(ni)}$ into a convolution, numerical evaluation of
(\ref{freeEni}) proves more challenging in the triatomic case because
of the complex, nine-dimensional form of $s$.

To address this, we observe here that $\Omega^{(ni)}$ for water can
\emph{also} be brought into a ``convolution form'' by making use of
the following mathematical identity, which is readily derived using
the addition theorem for spherical harmonics,
\begin{equation}
\delta(\hat{\mathbf{v}}\cdot\hat{\mathbf{w}}-\cos\zeta)=
\sum_{lm}2\pi P_{l}(\cos\zeta)Y^{*}_{lm}(\hat{\mathbf{v}})Y_{lm}(\hat{\mathbf{w}}),
\label{MathId}
\end{equation}
where $\hat{\mathbf{v}}$ and $\hat{\mathbf{w}}$ are unit vectors,
$\zeta$ is an arbitrary angle, $P_{l}$ is a Legendre polynomial and
the $Y_{lm}$'s denote the spherical harmonics.  (To prove
(\ref{MathId}), first expand the delta function as a sum of pairs of
Legendre polynomials on the interval $[-1;1]$ and then use the
spherical-harmonic ``addition theorem'' to express
$P_{l}(\hat{\mathbf{v}}\cdot\hat{\mathbf{w}})$ as a sum of pairs of
spherical harmonics.)

Substituting (\ref{MathId}) and (\ref{sWater}) into (\ref{freeEni})
results in a computationally efficient form for the non-interacting
free energy $\Omega^{(ni)}$,
\begin{eqnarray}
\Omega^{(ni)}=-k_{B}Tn_{l}\int
d^{3}r_{0}e^{-\beta\Psi_{0}(\mathbf{r}_{0})}\nonumber\\
\times\sum_{lm}4\pi P_{l}(\cos\theta_{B})f^{(1)}_{lm}(\mathbf{r}_{0})f^{(2)}_{lm}(\mathbf{r}_{0}),
\label{OmeganiConv}
\end{eqnarray}
where $f^{(1)}_{lm}$ and $f^{(2)}_{lm}$, given by
\begin{eqnarray}
f^{(1)}_{lm}(\mathbf{r}_{0})=\int d^{3}r_{1}
\frac{\delta(|\mathbf{r}_{1}-\mathbf{r}_{0}|-B)}{4\pi B^{2}}
Y^{*}_{lm}(\hat{\mathbf{r}}_{10})e^{-\beta\Psi_{1}(\mathbf{r}_{1})}\nonumber\\
f^{(2)}_{lm}(\mathbf{r}_{0})=\int d^{3}r_{2}
\frac{\delta(|\mathbf{r}_{2}-\mathbf{r}_{0}|-B)}{4\pi B^{2}}
Y_{lm}(\hat{\mathbf{r}}_{20})e^{-\beta\Psi_{2}(\mathbf{r}_{2})},
\end{eqnarray}
are now  convolutions and can be efficiently evaluated with fast-Fourier-transform techniques.
Transformation to Fourier space greatly simplifies the
above convolutions,
\begin{eqnarray}
f^{(1)}_{lm}(\mathbf{k})=(-i)^{l}j_{l}(|\mathbf{k}|B)Y^{*}_{lm}(\hat{\mathbf{k}})
{\mathcal F}[e^{-\beta\Psi_{1}(\mathbf{r})}]\nonumber\\
f^{(2)}_{lm}(\mathbf{k})=(-i)^{l}j_{l}(|\mathbf{k}|B)Y_{lm}(\hat{\mathbf{k}})
{\mathcal F}[e^{-\beta\Psi_{2}(\mathbf{r})}],
\label{FFTs}
\end{eqnarray}
where $j_{l}$ denotes the spherical Bessel functions of the first
kind and ${\mathcal F}[\exp(-\beta\Psi_{1}(\mathbf{r}))]$ denotes the Fourier
transform of $\exp(-\beta\Psi_{1}(\mathbf{r}))$.

To evaluate (\ref{OmeganiConv}) numerically, we first choose a maximum
value of $l$, $l_{max}$, after which we truncate the infinite sum. In
general, the choice of $l_{max}$ depends on the density profile that
has to be represented. We find in the capacitor calculation presented
below that the interaction site densities for water adjacent to the
capacitor wall are fairly smooth in the linear response regime and can
be sufficiently resolved with $l_{max}=10$. If strong external fields
are applied, sharp features in the density profile develop and we use
$l_{max}=40$ to be absolutely sure of a highly converged description.

\subsection{Intermolecular Free Energy}
First, we approximate $f^{ex}(n)$ as a sixth-order polynomial,
\begin{equation}
f^{ex}(n)=\sum_{p=0}^6 f_p n^{p},
\label{fex}
\end{equation}
and adjust its coefficients to reproduce the seven conditions
represented by (a) the thermodynamic stability of liquid and vapor
phases, (b) their coexistence, (c) the experimental liquid and vapor
densities, (d) the experimental bulk modulus of the liquid ($B_{l}$)
and (e) the derivative of the bulk modulus with respect to the
pressure $P$ at standard temperature and pressure
($T=25\;^\circ$C and $P=101.325\;\text{kPa}\equiv 1\;$atm).
Table~\ref{fexInput} summarizes the experimental input used to
determine the coefficients $f_p$, and Table~\ref{fexCoeffs} gives the
actual numerical values of the coefficients used in our calculations
below.  (Clearly, not all of the digits given in the table are
significant, we give them only to make our results numerically
reproducible.)  The values in the table are given in atomic units
($1~\text{hartree} \approx 27.21~\text{eV}$, $1~\text{bohr} \approx
0.5291$~\AA) as this is the system which our software employs.


\begin{table}
  \setlength{\doublerulesep}{0\doublerulesep}
  \setlength{\tabcolsep}{5\tabcolsep}
  \begin{tabular}{c c c c}
    \hline\hline\\
    $\rho_l$ & $\rho_v$ & $B_{l}$ & $\partial B_{l}/\partial P$\\
    $[\text{kg}/m^{3}]$ & $[\text{kg}/m^{3}]$ &
    $[\text{GPa}]$& (dimensionless) \\
    \hline\\
    997.1 & 0.023  & 2.187 & 5.8\\
    \hline\hline
  \end{tabular}
  \caption{Experimental inputs to construction of $f^{ex}$ from \cite{CRC} and \cite{Wilhelm}: 
    liquid density ($\rho_l$), vapor density ($\rho_v$), liquid bulk modulus ($B_l$) 
    and derivative of modulus with respect to pressure ($\partial B_{l}/\partial P$).}
  \label{fexInput}
\end{table}

\begin{table}
  \setlength{\doublerulesep}{0\doublerulesep}
  \setlength{\tabcolsep}{5\tabcolsep}
  \begin{tabular}{l l l}
    \hline\hline\\
    $f_{0}$ & $[\text{hartree}\times\text{bohr}^{-3}]$ & $\;\;\;3.7608\;5540\;7008\;46\times
    10^{-18}$\\
    $f_{1}$ & $[\text{hartree}]$ & $-9.9058\;0951\;0335\;88\times
    10^{-11}$\\
    $f_{2}$ & $[\text{hartree}\times\text{bohr}^{3}]$ & $\;\;\;8.6971\;6797\;7040\;56\times
    10^{-4}$\\
    $f_{3}$ & $[\text{hartree}\times\text{bohr}^{6}]$ & $-2.5454\;6838\;7756\;12\times
    10^{3}$\\
    $f_{4}$ & $[\text{hartree}\times\text{bohr}^{9}]$ & $\;\;\;8.9610\;5939\;0318\;49\times
    10^{5}$\\
    $f_{5}$ & $[\text{hartree}\times\text{bohr}^{12}]$ & $-1.2391\;1957\;0697\;09\times
    10^{8}$\\
    $f_{6}$ & $[\text{hartree}\times\text{bohr}^{15}]$ & $\;\;\;6.4173\;3045\;2123\;21\times
    10^{9}$\\
    \hline\hline
  \end{tabular}
  \caption{Coefficients of polynomial parametrization of $f^{ex}$
    given to all digits used in our software in order to ensure
    numerical reproducibility
    of our results.  Any individual coefficient contains at most two
    or three significant figures.}
  \label{fexCoeffs}
\end{table}

Next, we can determine the coefficients in the integrand in
(\ref{Fex}) from knowledge of the constant term in the long-wavelength
expansion of the Hessian $\partial^{2}_{\mathbf{n}}U$.  The entries in
this term (which is a matrix) are related to various material response
properties such as as the bulk modulus, but unfortunately, we do not
have access to data for all of the material properties.  In the
absence of data, we take this matrix to be proportional to its value for the
non-interacting case with a proportionality constant set to ensure the
correct bulk modulus.  This then fixes the
integrand in (\ref{Fex}) and allows it to be written entirely in terms
of rational numbers in the form,
\begin{eqnarray}
-\frac{19}{20}f^{ex}(\overline{n}_{0})+
\frac{3}{10}f^{ex}(\overline{n}_{1})+\frac{3}{10}f^{ex}(\overline{n}_{2})\nonumber\\
+\frac{27}{20}f^{ex}\left(
\frac{\overline{n}_{0}+\overline{n}_{1}+\overline{n}_{2}}{3}
\right).
\label{fexH2O}
\end{eqnarray} 

Next, to construct the matrix $K$ we use the partial charges of the
SPC/E water model \cite{BerendsenSPCE}, i.e.
$q_{1}=q_{2}=-q_{0}/2=0.4238\,e$ ($e$ being the proton charge) and the
fact that the dielectric constant of the liquid with intra-molecular
correlations {\em only} is $\epsilon^{(ni)}=(1-4\pi \beta
n_{l}p^{2}/3)^{-1}$, where $p$ is the dipole moment of an SPC/E water
molecule. Also, we employ the experimental value \cite{CRC},
$\epsilon=78.4$, for the macroscopic dielectric function of liquid
water at standard conditions described above.

The use of $K$ as leading order term is only justified for
\emph{long-wavelength} (small $\mathbf{k}$) properties of the liquid:
we therefore cut off its slowly decaying, large $\mathbf{k}$ tail by
multiplying it in Fourier space with a crossover function
$\lambda_{cr}$ given by
\begin{equation}
\lambda_{cr}(\mathbf{k})=\frac{1}{\left(1+|\mathbf{k}|/k_{c}\right)^{4}},
\end{equation}
with $k_{c}=0.33$~bohr, chosen to keep the $C$ functions as band-width
limited as possible.

\begin{figure}
\includegraphics[width=8.cm]{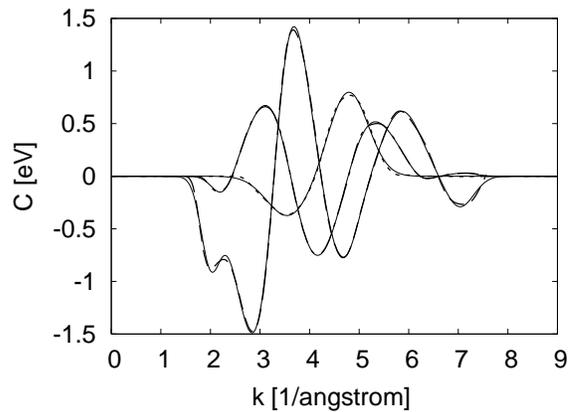}
\caption{Comparison of $C_{00}$ (thick dashed curve), $C_{01}$ (thick
  dash-dotted curve) and $C_{11}$ (thick dotted curve) and the
  corresponding parametrized forms (thin solid curves) given by
  (\ref{fits}) and Table~\ref{CfuncCoeffs}. Note that $C_{12}$ is not
  shown, since it would be visually indistinguishable from $C_{11}$.}
\label{Cfunctions}
\end{figure}

Then we determine $C$ as described above using the partial structure
factor data for the uniform liquid from Figure~1 in \cite{SoperWater}.
The best, explicit partial structure factor data which we have found
are measured at standard pressure, but at a temperature
$T=20\pm3\;^\circ$C, which differs slightly from the standard
temperature ($T=25\;^\circ$C) employed in this work. Comparing
experimental radial distribution functions at various temperatures
\cite{SoperTemp}, we find that the differences remain smaller than
10\% over a temperature range of 30 Kelvin. We therefore expect an
overall error of no more than one or two percent due to this
difference in temperatures.  Next, adjusting the smoothing parameter
to give the experimental surface tension of $71.98\times
10^{-5}\;$N/m\cite{CRC}, measured at standard conditions, yields
$r_{0}=4.2027~\text{bohr}$ (to all the digits used in our software).
Finally, to provide a computationally transferable representation of
$C$, we parametrize each component $C_{\alpha\gamma}$ by a sum of
Gaussians in Fourier space
\begin{equation}
C_{\alpha\gamma}(\mathbf{k})=\sum_{i=1}^{N_{\alpha\gamma}}A^{(i)}_{\alpha\gamma}\exp
\left\{-B^{(i)}_{\alpha\gamma}\left(|\mathbf{k}|-C^{(i)}_{\alpha\gamma}\right)^{2}\right\}.
\label{fits}
\end{equation}
Table~\ref{CfuncCoeffs} summarizes the resulting fitting coefficients
(to all decimal places used in our software).  To reproduce the full
matrix, note that $C$ is symmetric and that the identity of the
hydrogens imposes $C_{11}=C_{22}$.  Figure~\ref{Cfunctions} compares
our extracted $C$ functions to the resulting numerical fits, the
latter of which are used in all of the calculations presented below.

\begin{table}
  \setlength{\doublerulesep}{0\doublerulesep}
  \setlength{\tabcolsep}{5\tabcolsep}
  \begin{tabular}{l l l l l}
    \hline\hline\\
    & \multirow{2}{*}{$i$}& $A^{(i)}_{\alpha\gamma}$&
    $B^{(i)}_{\alpha\gamma}$ &
    $C^{(i)}_{\alpha\gamma}$ \\
    & & $[\text{hartree}]$ & $[\text{bohr}^{2}]$ & $[\text{bohr}^{-1}]$\\
    \hline\\
  \multirow{6}{*}{$C_{00}$}& $1$ & $-0.0271\;1530$     & $1.0569\;0000$  & $0.0204\;8180$\\
          & $2$ & $-0.0795\;5700$      & $1.6385\;3000$ & $0.1295\;1300$\\
          & $3$ & $\;\;\; 0.0966\;4800$  & $1.8781\;5000$ & $0.0736\;7760$\\
          & $4$ & $-0.0291\;5170$     & $2.4678\;7000$ & $0.0563\;6360$\\
          & $5$ & $\;\;\; 0.0227\;0520$ & $3.1004\;7000$ & $0.0844\;1320$\\
          & $6$ & $-0.0109\;0780$     & $3.7162\;0000$  & $0.0452\;0230$\\
    \hline\\
   \multirow{6}{*}{$C_{01}$}& $1$ & $-0.0080\;1259$    & $1.1985\;9000$ & $0.0280\;2800$\\
          & $2$ & $\;\;\; 0.0379\;8360$ & $1.7289\;7000$ & $0.1175\;3400$\\
          & $3$ & $-0.0380\;7370$     & $2.1583\;9000$ & $0.2133\;3600$\\
          & $4$ & $\;\;\; 0.0254\;5760$ & $2.7211\;2000$ & $0.1541\;5700$\\
          & $5$ & $-0.0029\;1329$    & $3.2989\;4000$ & $0.0265\;2820$\\
          & $6$ & $\;\;\; 0.0010\;9967$& $3.7659\;0000$  & $0.0315\;5180$\\
    \hline\\
  \multirow{2}{*}{$C_{11}$}& $1$ & $-0.0139\;5900$      & $1.8869\;7000$ & $0.1041\;8600$\\
          & $2$ & $\;\;\; 0.0295\;7760$ & $2.5316\;4000$ & $0.0869\;8480$\\
    \hline\\
  \multirow{2}{*}{$C_{12}$}& $1$ & $-0.0139\;5900$      & $1.8869\;7000$ & $0.1041\;8600$\\
          & $2$ & $\;\;\; 0.0295\;7760$ & $2.5316\;4000$ & $0.0869\;8480$\\
  \hline\hline
  \end{tabular}
  \caption{Coefficients of Gaussian parametrized C functions.  Note
    that, by symmetry, $C_{\alpha\gamma}=C_{\gamma\alpha}$ and
    $C_{11}=C_{22}$. ($C_{12}$ and $C_{11}$ turned out to be
    numerically indistinguishable.) Coefficients are given to all
    digits used in our software to ensure numerical reproducibility of
    our results.  The index $0$ refers to the oxygen site, the indices
    $1$ and $2$ refer to the two hydrogen sites.}
  \label{CfuncCoeffs}
\end{table}

\subsection{Recipe for Evaluating the Free Energy Functional for Water}

In sum, to compute the free-energy functional for a given set of
effective potentials $\Psi_\alpha(\mathbf{r})$, one first evaluates
the intra-molecular contribution via (\ref{OmeganiConv}) and
(\ref{FFTs}).  Next, one determines the site densities via
(\ref{eq:getns}), which allows one to evaluate the second term in
(\ref{FullOmega}) trivially.  Finally, one evaluates the last term
in (\ref{FullOmega}) according to (\ref{U}), with $K$ given by
(\ref{K}), $C$ parameterized by the coefficients in
Table~\ref{CfuncCoeffs}, and $F^{ex}$ computed according to
(\ref{Fex}), (\ref{fex}), (\ref{fexH2O}).

\section{Water in a Parallel Plate Capacitor}
To demonstrate the ability of our density-functional theory to
describe water in an inhomogeneous environment and to capture the
dielectric response of the liquid, we study the behavior of the liquid
in a parallel plate capacitor.  We carry out our study at standard
temperature and pressure described above, taking the system to be
homogeneous in the two dimensions parallel to the capacitor plates.
Along the perpendicular direction, we impose periodic boundary
conditions, defining a unit cell containing \emph{two} parallel plate
capacitors with opposing externally applied electric fields.  This
arrangement makes the electrostatic potential a periodic function and
eliminates undesired electrostatic interactions between the
capacitors.  Each capacitor has a plate separation of
$106$~\AA~(200~bohr) and the total length of the cell is
$423$~\AA~(800~bohr).

In our study we consider both purely repulsive, ``hard-wall'' plates
as well as ``attractive'' plates with a region of attraction near the
plates.  For the hard-wall case, the plates are essentially infinite
potential hard walls acting on both the oxygen and hydrogen sites of
the water molecules.  As a practical numerical matter, to reduce
aliasing effects in the use of the numerical Fourier transform we
approximate such walls by purely repulsive Gaussian potentials of
sufficiently narrow width and large amplitude so that the results
below are insensitive to the width and amplitude of the Gaussian
employed.  In particular, we use
\begin{equation}
\phi_{hw}(z)=A\exp\left(-\frac{z^{2}}{2a^{2}}\right),
\end{equation}
with $A=150~k_{B}T$ and $a=0.5$~\AA.  For the attractive case, we
employ a ``9-3'' (planarly integrated Lennard-Jones) potential
acting on the oxygens {\em only}, specifically,
\begin{equation}
\phi_{at}(z)=C\left[\left(\frac{D}{z+z_{\text{shift}}}\right)^{9}-
\left(\frac{D}{z+z_{\text{shift}}}\right)^{3}\right]
\end{equation}
with $C=9\times 0.15/(2\sqrt{3})$~eV, $D=\frac{5}{3^{1/6}}$~\AA~and
$z_{\text{shift}}=2$~\AA.  This potential has its minimum at a
distance of 3~\AA~ from the wall at a depth $0.15$~eV.  We choose this
depth as typical of the interactions in such systems.  For instance,
Berkowitz and co-workers \cite{BerkowitzWallPot} employ a corrugated
potential with an average depth of about 0.40~eV, which is also in the
range of a couple of tenths of an electron Volt.

To determine the thermodynamic state of the system we then expand the
density fields on a real space grid with 8192 sampling points ($\sim20$
points/\AA), and minimize the grand free energy (\ref{FullOmega})
using standard numerical conjugate gradient techniques without any
preconditioning.  

\subsection{Results}

\begin{figure}
\includegraphics[width=8.cm]{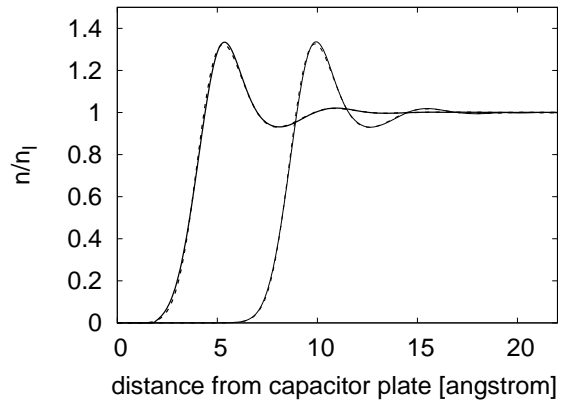}
\caption{Hydrogen- (thick dotted curve) and oxygen- (thick solid
curve) site density in weak field and zero field (thin curves) versus
distance from hard-wall plate.  (Note that the hydrogen and oxygen
densities are nearly visually indistinguishable.)}
\label{ZeroField}
\end{figure}

\subsubsection{Hard-wall plates}

We begin with a brief discussion of our results for the hard wall case
as an idealized reference for comparison.  Figure~\ref{ZeroField}
shows our results for oxygen and hydrogen equilibrium density profiles
in the absence of an externally applied electrostatic displacement
($D=0$) and compares them to the density profiles of hydrogen and
oxygen sites in a relatively weak external field ($\alpha \equiv \beta
p D/\epsilon \approx 0.009$).  The zero-field profiles exhibit an
extended gas phase region up to a distance of $\sim 10$~\AA~from the
plates. This ``lingering'' gas phase region exists because molecules
can, at very little free-energy cost, minimize the influence of the
repulsive plates by leaving the system.  (Recall that
(\ref{FullOmega}) is written in the grand canonical ensemble, so that
the system can exchange particles with an external reservoir.)  As
soon as a relatively weak external electric displacement is applied,
however, it becomes favorable for the dipolar molecules to enter the
system because they can lower their energy by partially aligning with
the field within the capacitor.  For even quite small fields, this
effect causes the capacitor to fill with liquid, thereby eliminating
most of the gas phase region.  The result of this is then an almost
rigid shift of the density profiles toward the capacitor plates, with
an appearance and general shape which remains nearly fixed over a
large range of fields.

Considering the fine details of the density profiles, we note that,
contrary to our previous results for diatomic liquids, a relatively
small (barely noticeable in the figure) asymmetry exists at zero
external field between the oxygen and hydrogen site densities so that
there is a non-vanishing electrostatic charge density in the vicinity
of the capacitor plates --- even in the absence of an externally
applied electric field.  (Note: because the oxygen site carries a
charge which is double in magnitude compared to the hydrogen site and
because there are two independent but identical hydrogen sites, the
magnitude of the charges carried by the oxygen and hydrogen density
fields are directly proportional to the curves in the figure, {\em
  with the same constant of proportionality} for each.)  By symmetry,
the net charge which the plates induce must integrate to zero;
however, a finite surface dipole remains with a corresponding
potential jump which we find in the zero-field hard-wall case to be
+$0.041$~V across the solid-liquid interface, with the potential being
higher in the solid.

\begin{figure}
\includegraphics[width=8.cm]{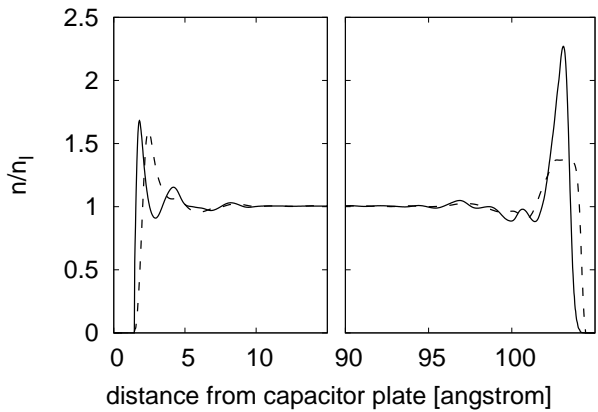}
\caption{Hydrogen- (dashed curve) and oxygen- (solid curve) site
density in strong field versus distance from hard-wall plate.}
\label{D197}
\end{figure}

Figure~\ref{D197} shows the equilibrium density profiles in a
relatively strong ($\alpha\approx 0.25$) external field directed from
left to right in the figure.  The relative abundance of oxygen sites
close to the left plate (which carries a positive external charge to
generate the field) and of hydrogen sites next to the right plate
(which carries a negative external charge) indicates that the water
molecules exhibit a strong tendency to align their dipoles with the
applied field.  Interestingly, this leads to quite different density
profiles on the opposite sides of the capacitor.  On the left
(positive) plate, the first peak in the oxygen density is followed by
a peak in the hydrogen density of similar shape, width and height.
This we interpret to be a direct consequence of alignment of the
molecules with the field and the rigid intra-molecular bonding between
hydrogen and oxygen atoms, an effect quite similar to what was
observed in our previous study of liquid hydrogen
chloride\cite{JLTAA}. The distance between these peaks, 0.65~\AA, is
within about 10\% of 0.58~\AA, what one expects from the basic
geometry of the water molecule assuming that the molecules fully align
with the field.

On the right (negative) plate, we again find a well-defined peak in
the oxygen density, though the peak is somewhat stronger than on the
left.  We find also that the hydrogen density, rather than being also
sharply peaked, is now spread significantly around the oxygen peak.
This case is different, because the hydrogen sites, now pointing away
from the liquid region, cannot form hydrogen bonds if the molecules
align fully with the field.

\subsubsection{Attractive plates}

\begin{figure}
\includegraphics[width=8.cm]{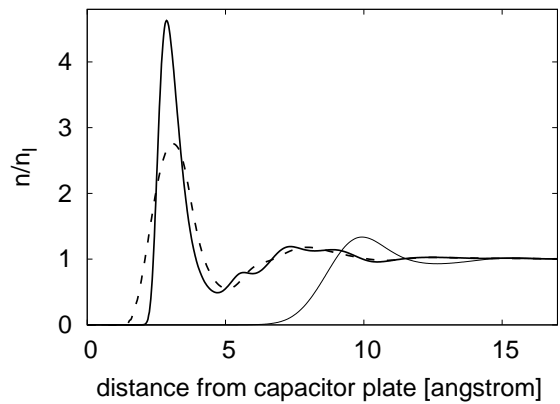}
\caption{Hydrogen- (thick dashed curve) and oxygen- (thick solid
curve) site density in strong field versus distance from attractive
wall. Reference result for the zero field oxygen density at repulsive
wall (thin solid curve).}
\label{LJZeroField}
\end{figure}

Figure~\ref{LJZeroField} compares the zero-field results for the
hard-wall plates with those of the attractive plates.  There is no
longer a ``lingering'' gas phase region because the minimum in the
interaction potential (acting only on the oxygen sites in this case)
with the plates sets the location of the main peak in the oxygen
density.  This minimum, as described above, occurs at a distance of
about 3~\AA~from the attractive walls and corresponds well with the
peaks in the oxygen and hydrogen density profiles. The height of the
oxygen peak is larger than the density in the bulk by a factor of
$\sim 4.7$, which is similar to the results of molecular dynamics
calculations of water adjacent to platinum surfaces
\cite{BerkowitzWallPot} and experimental findings \cite{Toney}.
Finally, there is also a net dipole layer in our attractive case
leading now to a significantly stronger potential jump at the
liquid-solid interface of $+0.220$~V, with the potential in the solid
again being higher.
This value is typical of realistic systems: from
molecular dynamics data of water adjacent to a platinum
surface\cite{BerkowitzDielectric}, we extract a potential jump of
$\sim +0.5$~V with the potential being higher in the solid.
\begin{figure}
\includegraphics[width=8.cm]{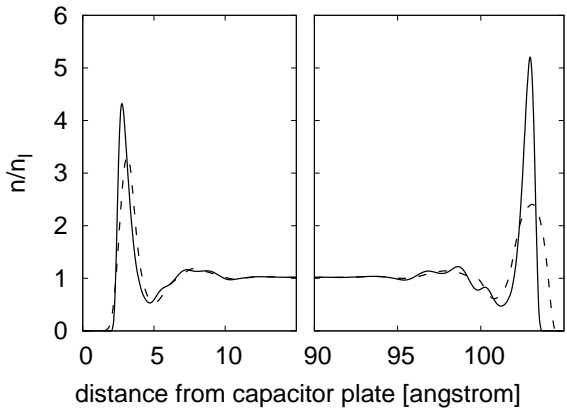}
\caption{Hydrogen- (dashed curve) and oxygen- (solid
curve) site density in strong field versus distance from attractive
wall.}
\label{PhilD197}
\end{figure}

When a strong field (again, $\alpha\approx 0.25$) is applied to the
capacitor with attractive walls (Figure~\ref{PhilD197}), we find a
rearrangement of the liquid structure which exhibits various
quantitative differences compared to the hard-wall case: the oxygen
peaks are now significantly more pronounced (reaching values near 4
and 5 times the equilibrium liquid density, as opposed to the previous
peak values near 1.5 and 2.2) as the oxygen atoms settle into the
attractive potential wells near the walls.  Also, the peak near the
positive plate is also now notably broader (with a width of 1.6~\AA~ at
the value of the equilibrium, as opposed to the previous value of
1.0~\AA). Finally, the hydrogen peak is significantly less pronounced
relative to the oxygen peak and the peak position is now less
displaced (0.4~\AA) from the oxygen peak position.  Despite these
quantitative differences, however, the qualitative features at strong
fields are quite similar in the attractive and the hard-wall cases.
Namely, on the left (positive) plate we find a well-defined peak in
the oxygen density followed by a corresponding peak in the hydrogen
density, and on the right (negative) plate there is a well-defined
oxygen peak which is surrounded by a more diffuse peak in the hydrogen
density.  Also, as with the hard wall case, we find that the hydrogen
peak is off-center from the oxygen peak and displaced toward the
negative plate to which it is attracted.

This suggests that, at these large field strengths, general
features are independent of the exact form of the liquid-solid
interaction.  In fact, we note that this sort of asymmetry between
positive and negative plates is also evident in molecular dynamics
calculations which, at fields near $\alpha=0.25$, also tend to show
that, near the positive plate, the oxygen density exhibits peaks which
are followed by peaks of similar appearance in the hydrogen density,
whereas, near the negative plate, there is a well-defined peak in the
oxygen density but that the hydrogen density exhibits either a broader
peak or multiple sub-peaks in the vicinity of the main oxygen
peak\cite{BerkowitzDielectric, BerkowitzPRL}.
\begin{figure}
\includegraphics[width=8.cm]{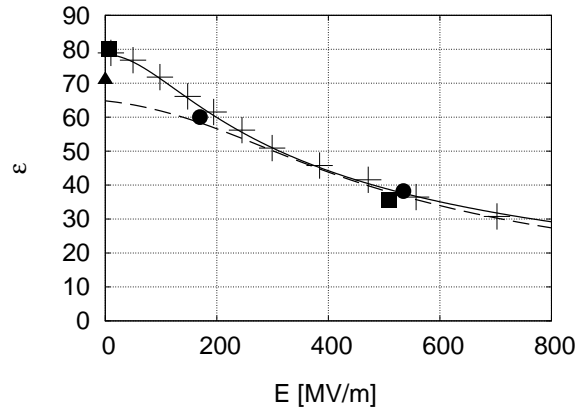}
\caption{Dielectric function versus total field: density-functional
  results (crosses), non-linear electrostatics (solid curve),
  analytical result of \cite{Booth}(dashed line), molecular
  dynamics results of \cite{BerkowitzDielectric}(circles),
  \cite{Sutmann}(squares) and \cite{Kusalik}(triangles).}
\label{Dielectric}
\end{figure}

\subsubsection{Nonlinear dielectric response}

The equilibrium site densities also determine the dielectric function
of water $\epsilon(D)=D/E(D)$ with $E$ being the total electric field
given by the sum of the electric displacement and the induced electric
field. Although $E$ has a complicated form close to the walls, it
rapidly approaches the constant value $E=D-4\pi P$ as the distance
from the walls increases, where $P$ is the induced polarization.  In
the present one-dimensional geometry, $P$ is proportional to the
induced surface charge which appears in the equilibrium density
profiles.

Figure~\ref{Dielectric} shows our calculated results for $\epsilon$
plotted as a function of $E$ in units of MV/m (equivalent to 1~mV/nm
or 0.1~mV/\AA) to simplify comparison with the results of previous
studies.  The figure shows that the results from our
density-functional theory are well-described by self-consistently
screened nonlinear electrostatics (solid curve), which gives the
polarization $P$ as the self-consistent solution to the equation
$P=P^{(ni)}(D-a_{\epsilon}4\pi P)$, where $P^{(ni)}(E)$ is the
response of a gas of noninteracting dipoles in a local field $E$ and
$a_{\epsilon}\equiv
\epsilon/(\epsilon-1)-\epsilon^{(ni)}/(\epsilon^{(ni)}-1)$ ensures
that the correct linear response is recovered when $D$ is weak.  The
solid curve reproduces our data well and so represents a convenient
analytic guide to the eye for our results.

Figure~\ref{Dielectric} compares our results to Booth's analytical
result for the nonlinear dielectric response of water \cite{Booth}.
This result (dashed curve in the figure) was obtained by extending
Onsager's and Kirkwood's theories of polar dielectrics to high field
strengths \cite{Booth} and also reproduces quite well explicit molecular
dynamics results for high fields.  It fails, however, at
small fields, eventually giving an incorrect linear dielectric
constant of about $\sim 65$.  We find very good agreement with this
established result for fields larger than $300$~MV/m.  We also compare
our results to various molecular dynamics calculations
\cite{BerkowitzDielectric, Sutmann, Kusalik} and find very good
agreement, except at zero field, where the results of the molecular
dynamics calculations often show a wide range of results, some
differing significantly from the experimental dielectric constant of
$\sim 80$ depending on the model potentials and numerical methods
employed\cite{SvichevScience}. The zero-field dielectric constant,
we, of course, reproduce exactly by construction.  Our results,
therefore, appear to be reliable over a broad range of applied fields.

\section{Summary and Conclusions}
This work demonstrates how to apply a new, general Kohn-Sham classical
density-functional approach for molecular liquids to what is arguably
the liquid of greatest scientific interest, water.  The resulting
theory has the quite tractable computational cost of the problem of a
noninteracting gas of molecules in a self-consistent external
potential and gives an exact account of the entropy associated with
the alignment of the molecular dipole moment with an external field.

After constructing the functional and giving a complete numerical
parameterization of all quantities so that our results are numerically
reproducible, we demonstrate the tractability of the approach by
applying it to the behavior of water in parallel plate capacitors,
with either hard-wall or attractive plates, over a range of applied
fields through which the dielectric constant $\epsilon$ of water
varies by over a factor of two.  We provide results for the
distribution of oxygen and hydrogen atoms near the capacitor plates
and find an asymmetry between the response at negatively and
positively charged plates in general qualitative agreement with
molecular dynamics results.  From the resulting charge distributions,
we extract predictions of the dielectric constant of water over a wide
range of field strengths (0 to 800~MV/m) and find results in very good
agreement with both previous theoretical and molecular
dynamics calculations.  The aforementioned exact treatment of the
entropy cost of aligning the molecules with the field is critical in
producing this accurate treatment of dielectric saturation effects.

With a successful framework in place, there is now a firm basis for
future improvements.  Given that we now have a good treatment of the
internal structure of the water molecule, the primary area remaining
for improvement is our choice to employ for this first work a somewhat
simplistic weighted-density functional form in (\ref{Fex}).  The
direct route to improve upon this is to leverage much of the existent
work in simple fluids, both in traditional weighted density-functional
theories\cite{CurtinAshcroft} and in fundamental measure
theory\cite{RosenfeldFundMeasTh}.  Another clear area for improvement
would be the inclusion of additional, ionic species into the
functional to allow for screening by electrolytes, which should be
relatively straightforward.

In sum, we present a new framework for a computationally tractable
continuum description of water which captures, within a unified
approach with a firm theoretical foundation, molecular-scale
correlations, entropic effects, microscopic dielectric screening,
surface tension, bulk thermodynamic properties, and dielectric
saturation.  This description thus includes {\em a priori} all of the
various effects which are generally added {\em a posteriori} to
continuum models of water.
Although less detailed than current explicit molecular treatments of
the solvent, freed from the need for computing thermodynamic averages,
a sufficiently accurate continuum approach such as we present has the
potential to open up much larger and complex systems to theoretical
study.
As such, the new framework is an ideal starting point for a variety of
new investigations into a wide class of subjects, which include
protein folding, molecular motors, membrane physics, drug design, the
electrochemical interface, liquid-phase catalysis and fuel cells.

\section{Acknowledgements}
T.A.A. was supported in part by NSF Grant No. CHE-0113670, and J.L. 
acknowledges support by DOE Grant No. DE-FG02-07ER46432.

\end{document}